\documentclass[12pt]{article}
\usepackage{graphicx}
\begin{document}

\title{High-Performance {\it Small-Scale}  Simulation of Star Clusters Evolution on Cray XD1}

\author{Keigo Nitadori
\thanks{
	Department of Astronomy, 
	School of Science, University of Tokyo, 
	Tokyo 113-0033, 
	\newline Email: nitadori@astron.s.u-tokyo.ac.jp},
Junichiro Makino
\thanks{
	Department of Astronomy, 
	School of Science, University of Tokyo, 
	Tokyo 113-0033, 
	\newline Email: makino@astron.s.u-tokyo.ac.jp}
and
George Abe
\thanks{
	Sony Corp. InFormation tech. Labs., 
	Tokyo 114-0001, Email: George.Abe@jp.sony.com}
}
\date{}

\maketitle

\begin{abstract}
In this paper, we describe the performance of an $N$-body simulation
of star cluster with 64k stars on a Cray XD1 system with 400 dual-core
Opteron processors. A number of astrophysical $N$-body simulations
were reported in SCxy conferences. All previous entries for
Gordon-Bell prizes used at least 700k particles. The reason for this
preference of  large numbers of particles is the parallel efficiency. It
is very difficult to achieve high performance on large parallel
machines, if the number of particles is small. However, for many
scientifically important problems the calculation cost scales as
$O(N^{3.3})$, and it is very important to use large machines for
relatively small number of particles. We achieved 2.03 Tflops, or
57.7\% of the theoretical peak performance, using a direct $O(N^2)$
calculation with the individual timestep algorithm, on 64k
particles. The best efficiency previously reported on similar
calculation with 64K or smaller number of particles is 12\%
(9 Gflops) on Cray T3E-600 with 128 processors.
Our implementation is based on highly scalable two-dimensional
parallelization scheme, and low-latency communication network of Cray
XD1 turned out to be essential to achieve this level of performance.

\end{abstract}

\section{Introduction}

Many astronomical objects, such as the Solar system, star clusters,
galaxies, and clusters of galaxies and the large scale structure of
the universe, are expressed as the system of many ``particles''
interacting through the Newtonian gravity. The numerical
simulation of such systems, usually called ``Gravitational $N$-body
simulations'', proved itself to be one of the most useful tools in
computational astrophysics. It has been used to study systems of all
scales, from the formation of the Moon \cite{KMI99} to the
entire visible universe \cite{MilleniumRun}.

Many of these simulations are computationally demanding, because of
three reasons. The first reason is that the calculation cost per
timestep, at least with naive implementation, scales as $O(N^2)$,
where $N$ is the number of particles in the system.

The second reason is that in many cases the evolution timescale of the
system is many orders of magnitude longer than the orbital timescale
of typical particles in the system. An example is the long term
simulation of the Solar system. We need to follow the orbits of
planets and other objects for $4.5\times 10^9$ years, and we need
around 100 timesteps per year or more to achieve reasonable
accuracy. In the case of the simulation of the formation of planets,
we need to follow the evolution of planetesimals for several million
orbits. This requirement of large number of orbits limits the number of
particles we can handle to around 1,000 or less 
\cite{KokuboIda02}, 
even on special-purpose computers such as GRAPE-4 \cite{Grape4} or
GRAPE-6 \cite{Grape6}. The same is true for the simulation of star clusters. Here,
the evolution timescale depends linearly on the number of particles,
and the number of timesteps per orbits increases as $O(N^{1/3})$, since
the distance which one particle can move in a single timestep must be
a small fraction of the distance to its nearest neighbor. In total,
the calculation cost increases as $O(N^{3.3})$.

The third reason is that different particles can have very different
orbital timescales. Since the gravitational force is an attracting
force, two particles can become arbitrary close to each other. Thus, the timestep need
to be adaptive. Moreover, it is not enough to just change the
timestep. If we use the variable, but single, timestep to integrate
all particles, when we increase $N$ the average step-size for star
cluster simulation shrinks as $1/N$ or somewhat faster. Thus, the
calculation cost becomes $O(N^4)$ instead of $O(N^{3.3})$.

We can solve the third problem by allowing individual particles to
have their own time and timestep \cite{Aarseth63, Aarseth85}. 
However, with this scheme the
number of particles we can integrate in parallel becomes much smaller
than $N$, and it becomes more difficult to use large scale parallel
computers even when $N$ is not very small.

Dorband et al. \cite{DHM03} reported the performance of parallel
implementation of this individual timestep (or block timestep)
algorithm, combined with simple $O(N^2)$ direct force calculation, on
a Cray T3E. They have fine-tuned the communication scheme so that
there will be maximal overlap between the communication and
calculation. Even so, if the initial distribution of particles is
highly anisotropic, they found that performance did not increase for
more than 64 processors, for $N=64{\rm k}$. More recent works on PC
clusters \cite{Gualandris2004} generally gave much worse scaling,
because the communication performance of PC clusters, when normalized
by the calculation speed, is orders of magnitude worse than that of
Cray T3E.

The use of special-purpose GRAPE processors does improve the
performance of single-node calculation quite significantly and made
it possible to perform long simulations with relatively small number
of particles such as 64K or less. However, in this case the overall
performance was limited by the speed of the single host computer,
since the parallelization over multiple GRAPE nodes involves much
bigger relative overhead compared to that of PC clusters, and it
turned out to be difficult to achieve any gain unless the number of
particles is quite large \cite{Grape6A}. 

In this paper, we report the result of the first calculation which used
nearly 1,000 processors for the simulation of star clusters with just
64K stars and achieved more than 50\% of the theoretical peak speed of
the machine. Average speed measured on a 400 CPU Cray XD1 (2.2GHz
Dual-Core Opteron) was 2.03 Tflops, while the nominal theoretical peak
speed is 3.52 Tflops.  Previous works, even when the parallel efficiency was
good, achieved less than 15\% of the theoretical peak speed of the
machine. Thus, our implementation is remarkable in two ways: it
is highly scalable for small number of particles, and it can achieve
quite high absolute performance.

In the rest of this paper, we describe how we achieved the good
scalability and high absolute peak performance at the same time. In
section 2 we outline the individual (block) timestep algorithm. In
section 3 we describe the two-dimensional parallelization scheme we
used. It was originally proposed by Makino\cite{Makino2002}. 
We describe the
modifications we made to achieve high performance. In section 4 we
briefly describe the highly optimized force calculation loop. The
details are given in Nitadori et al.\cite{NMH06}. 
In section 5 we describe
the result of a long calculation of a two-component star cluster with
64K particles. Section 6 sums up.

\section{Block timestep algorithm}

The basic idea of the individual timestep algorithm is to assign each
particle its own time and timestep. Thus, particle $i$ has its current
time $t_i$ and timestep $\Delta t_i$. The calculation proceeds in
the following way:

\begin{enumerate}

\item Select a particle with minimum $t_i + \Delta t_i$
\item Integrate that particle to its new time, $t_{i,new}= t_i +
\Delta t_i$ and determine its new timestep.
\item Go back to step 1.
\end{enumerate}

In order to integrate particle $i$, it is necessary to calculate the
gravitational forces from all other particles at $t=t_i + \Delta
t_i$. For all particles other than particle $i$, it is guaranteed that
$t_j \le t \le t_j + \Delta t_j$. To calculate the force on particle
$i$ at time $t$, we perform ``prediction'' of positions of other
particles at time $t$. In order to be able to do this prediction, we
use linear multistep method with variable stepsize as the integration
scheme.

A practical problem with this individual timestep algorithm is that only
one particle is integrated at each time. Thus, there is little room
for parallelization. McMillan \cite{McM86} introduced the so-called
blockstep algorithm, in which timesteps are quantized to integer
powers of two, in such a way that particles with the same timestep
also has the same time and can be integrated in parallel. This means
that the timestep should be chosen so that the current time is an
exact integer multiple of the timestep. This simple criterion gives us
the maximum possible parallelism \cite{Makino91b}.

Until early 1990s, the 4th-order variable-stepsize
Adams-Bashforth-Moulton linear multistep scheme adopted to
second-order differential equation had been used as the integration
scheme. Currently, 4th-order Hermite scheme 
\cite{Makino91a, MakinoAarseth1992}
is used. It has several advantages over the original ABM
scheme.

\section{Two-dimensional parallel algorithm with optimal load balance}

\subsection{One-dimensional parallelization}

Existing parallel implementation of
the block timestep algorithm \cite{DHM03,Grape6A,Gualandris2004} are
all based on what we call one-dimensional
parallelization, which divide the calculation of gravitational
interaction along one loop index. The basic force calculation loop,
for the block timestep algorithm, would look like:
\begin{verbatim}
for(ii=0; ii<nblock; ii++){
    i = list_of_particles_in_current_block[ii];
    clear the force on particle i;
    for(j=0; j<n; j++){
         if(j!=i){
              calculate and accumulate the force from j to i;
         }
     }
}
\end{verbatim}
Here, {\tt list\_of\_particles\_in\_current\_block} is the list of
particles to be integrated at the current block timestep. Existing
parallel implementations are one-dimensional, in the sense that the
parallelization is done either to the inner loop or the outer loop,
but not both.

One might think the one-dimensional parallelization is sufficient, if the
number of particles, $N$, is significantly larger than the number of
processors, $p$. However, that is not the case for the blockstep
algorithm, since the number of particles in the current block, $N_b$,
is much smaller than $N$. In a 64k-particle run which will be
described in section 5, on average $N_b$ is around 400. Thus,
the parallelization over the outer loop (what is sometimes called
$i$-parallelization) is inadequate for more than a few hundred
processors.  Parallelization of the inner loop ($j$-parallelization)
has its own problem. It requires summation over all processors, which
can be very slow on distributed memory machine with relatively slow
network.

Moreover, both $i$- and $j$-parallelization on a distributed-memory
machine requires $O(N_b)$ communication per blockstep per node,
independent of the number of nodes. The calculation per node is
$O(NN_b/p)$. Thus, maximum number of nodes with which we can achieve
reasonable parallel efficiency is $O(N)$. If the communication is
relatively slow, the coefficient in front of $N$ can be very small,
and that is the reason why the scalability was rather bad in previous
works.

\subsection{Basic two-dimensional scheme}

We can improve the scalability by applying $i$- and
$j$-parallelization at the same time. In the following, we summarize
the description by Makino \cite{Makino2002}. The "regular-bases"
Hyper-systolic algorithm\cite{HyperSys} is essentially the same. One
can argue that the basic idea of these schemes is to use up to $O(N^2)$
processors for $N$-body problem. Hillis and
Barnes\cite{HillisBarnes1987} described such algorithm.

Consider the case in which $p$ processors are organized as 
an $r \times r$ two-dimensional array ($r^2=p$).  $N$ particles are
divided into $r$ subsets, and processor $q_{ij}$ has the
$i$-th and $j$-th subsets. The calculation proceeds as follows:

\begin{enumerate}

\item Each of processor $q_{ij}$ calculates the forces from subset $j$
to subset $i$.
\item Take summation over the processors in the same row. The result
is stored in processors at diagonal location, $q_{ii}$. They now have
the total forces on their subset $i$.

\item Processors $q_{ii}$ broadcast the summed forces to the column
direction.

\item All processors update the particles in their subset $i$, and
diagonal processors broadcast the updated particles to the row
direction, so that each processor receives updated $j$-th subsets.

\end{enumerate}

In practice, with the individual timestep algorithm we calculate the
forces only on the particles in the current block (which we call {\it
active particles}), and all communications in the above procedure
applies only to these active particles. Thus, in this algorithm, each
processor send/receive $O(N_b/r)$ data, instead of $O(N_b)$ data as in
the case of one-dimensional scheme.

\subsection{Improved 2D algorithm}

The basic two-dimensional parallelization, applied to the block
timestep, has two problems. One is that the load-balancing can be very
bad if the number of active particles is
small. At each timestep, processor $q_{ij}$ calculates the force from
$j$-th subset to active  particles in $i$-th subset. Thus,
fluctuations in the number of active particles directly
affect the load balance. This effect can be serious if number of nodes
in one dimension $r$ is large, since the node with maximum number of
particles in the block would determine the calculation time. Another
is that we can use only a square array of processors.

Here, we present an improved parallelization scheme, which solves both
of the above two problems. The basic idea is still the same: to apply
both the $i$- and $j$-parallelization. However, the communication pattern and
the data decomposition are completely different from those in the basic
scheme described in the previous section.

Consider the case in which $p$ processors are organized as an $a
\times b$ two-dimensional array. Communications occurs either row-wise
or column-wise. In the actual implementation with MPI, for these two
patterns, we construct  MPI communicators, to keep the program simple
and to take advantage of the MPI library functions for aggregate
communications. 

$N$
particles are divided into $b$ subsets, and all processors $q_{xj}$
have the $j$-th subsets. The calculation proceeds as follows

\begin{enumerate}

\item Each processor construct the list of active particles from its
subset. Here, the lists of processors $q_{xj}$ are identical for all
values of $x$.

\item Each processor horizontally broadcasts the length of its 
active list and the minimum timestep in its subset.
We use {\tt MPI\_Allgather} for this purpose. 
If other processor has a smaller timestep,
the active list is "inactivated" (none of particles in its subsets
are integrated).

\item Each processor divides its active list to $a$ subsets, and
processor $q_{ij}$ selects $i$-th subset as its active list.

\item Each processor horizontally broadcasts its active list, so that
all processors in the one row will share their lists of active
particles. We use {\tt MPI\_Allgatherv} for this purpose. 

\item Processor $q_{ij}$ calculates the force from its subset $j$ to the
particles in the combined list of active particles.

\item We take the summation of partial forces 
over the processors of each row, and store the results to the
processors of the original location. We  use {\tt
MPI\_Reduce\_Scatter} for this purpose. At this stage, each processor
has the total force on its active particles.

\item Each processor broadcast the forces on its active particles to
all other processors in the same column. We use {\tt MPI\_Allgatherv}
for this purpose. Now All processors $q_{xj}$ has the forces on the
active particles of subset $j$.

\item Each processor integrates the orbits of its active particles.
Here, all processors $q_{xj}$ do redundant operations.

\end{enumerate}

\begin{figure}[htpd]
\begin{center}
\includegraphics[width=0.75\hsize]{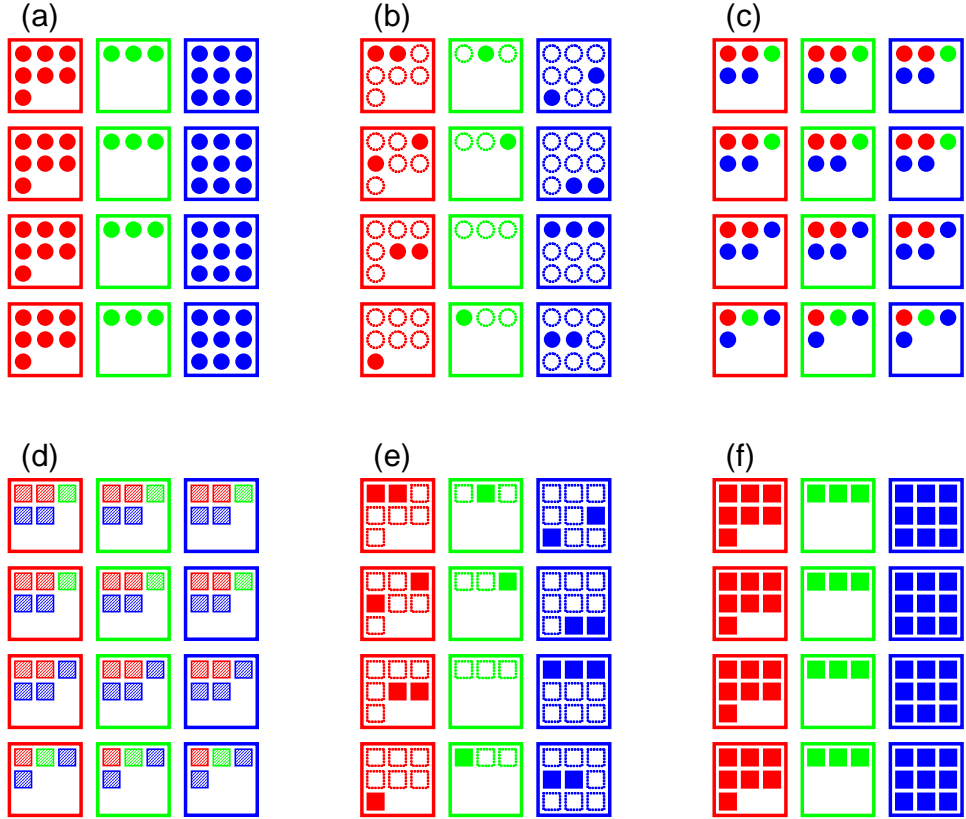}
\end{center}
\caption[The improved 2D algorithm]{
Improved 2D algorithm for processor array of $a=4, b=3$.
Circles denote positions, half-toned and filled square denote
partial and full forces, respectively.
(a)  7, 3 and 9 active particles are found in columns 1, 2 and 3.
(b) Each processor in one column decides which particles to handle.
(c) Horizontal broadcast of the positions.
(d) Force calculation
(e) Summation of the partial forces. Completed forces are send back to
the original locations. 
(f) Vertical broadcast of the forces.
}
\label{fig:2d-alg}
\end{figure}

Figure \ref{fig:2d-alg} illustrates how our new 2D algorithm works.
In this example, even if the number of active particles varies widely,
the amount of work on each processor is quite well balanced.

\section{Optimized force calculation loop}
\subsection{Functions and the number of operations}

In gravitational $N$-body simulations, if parallel efficiency is
reasonable, by definition almost all computing time is spent to
calculate the gravitational force from one particle to another. With
the Hermite integration scheme, what we need to calculate is the
acceleration, its time derivative, and the potential, given by

\begin{eqnarray}
\mathbf{a}_i & = &
\sum_j {m_j {\bf r}_{ij} \over (r_{ij}^2 + \varepsilon^2)^{3/2}} \\
%
{\bf j}_i & = & {{\bf{\dot a}}_i} \ =\ 
\sum_j m_j \left[{{\bf v}_{ij} \over (r_{ij}^2 + \varepsilon^2)^{3/2}}
- {3({\bf v}_{ij}\cdot{\bf r}_{ij}){\bf r}_{ij} \over (r_{ij}^2
+ \varepsilon^2)^{5/2}} \right] \\ 
 \phi_i & = &
-\sum_j {m_j \over (r_{ij}^2 + \varepsilon^2)^{1/2}}
\end{eqnarray}
where $\mathbf{a}_i$ and $\phi_i$ are the gravitational acceleration and
the potential of particle $i$, the jerk ${\bf j}_i$ is the time
derivative of the acceleration used for the Hermite integration \cite{MakinoAarseth1992}, 
and ${\bf r}_i$, ${\bf v}_i$, and $m_i$
are the position, velocity and mass of particle $i$, ${\bf r}_{ij} =
{\bf r}_{j} - {\bf r}_{i}$ and ${\bf v}_{ij} = {\bf v}_{j} - {\bf
v}_{i}$.

The calculation of $\mathbf{a}$ and $\phi$ requires nine multiplications,
10 addition/subtraction operations, one division and one square root
calculation.  Warren et al.  \cite{WSB97} used 38 as the total number
of floating point operations for this pairwise force calculation, and
that number has been used by number of researchers. So we follow that
convention.

In addition, it requires 11 multiplications and 11
addition/subtraction operations to calculate ${\bf j}$.  Thus, the
total number of floating-point operations per inner force loop for the
Hermite scheme is  60\cite{NMH06}.

\subsection{Performance tuning}

The full details of the optimization we applied is described in
\cite{NMH06}. The basic idea is to take full advantage of new
functionalities of SSE and SSE2 instructions, without sacrificing the
accuracy. The SSE2 instruction set offers the register-based
instructions which is easier to optimize than the stack-based x87
instruction set. It also gives SIMD operations for two
double-precision words. The SSE instruction set gives SIMD operation
on four single-precision words. In addition, it gives fast
approximation for inverse square root, which is extremely useful for
gravitational force calculation. Our optimized force calculation
function uses double-precision SSE2 instructions for the first
subtraction of positions and final accumulation of the calculated
force, but use SSE single-precision instructions for other
operations. Also, fast approximate inverse square root and one
Newton-Raphson  iteration is used to calculate the inverse square
root. 

On AMD K8 (Athlon 64 or Opteron) processors, our force loop calculates
four gravitational interactions in 120 CPU cycles. In other words,
it effectively performs 240 floating-point operations in 120 cycles.
The peak performance we can achieve for the force calculation on a
single core with 2.2 GHz clock speed is thus 4.4 Gflops. This number
happens to be exactly the same as the theoretical peak performance of
K8 processors for double-precision operations.

We used an Cray XD1 with 400 2.2GHz dual-core Opteron
processors. Therefore, the peak speed we can achieve for the force
calculation on single chip is 8.8 Gflops and that for the entire
machine is 3.52 Tflops.

\section{Simulation of star cluster evolution}

In this section, we report the performance of a
simulation of the thermal evolution of a star cluster
using the parallel individual timestep code we developed.

\begin{figure}
\begin{center}
	\includegraphics[width=\hsize]{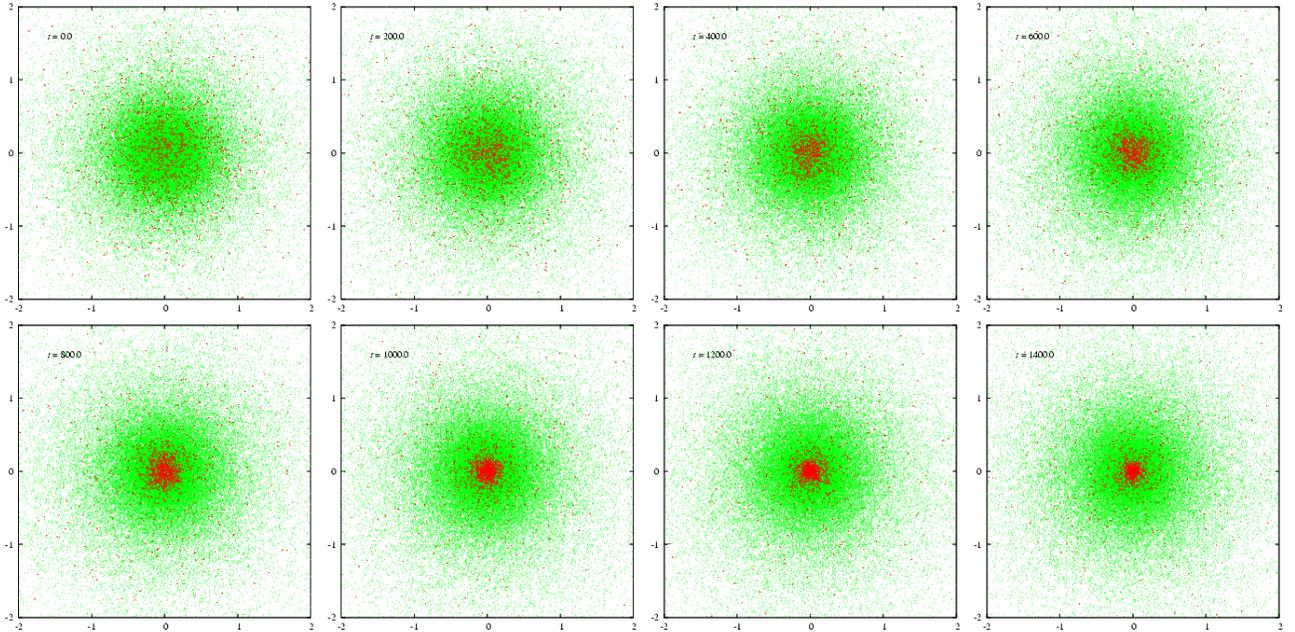} \\
\end{center}
\caption{
	The evolution of the star cluster.
	The light particles are plotted in green and the heavy particles
	are plotted in red.
}
\label{fig:evolve}
\end{figure}

\begin{figure}
\begin{center}
  \includegraphics[height=0.75\hsize, angle=270]{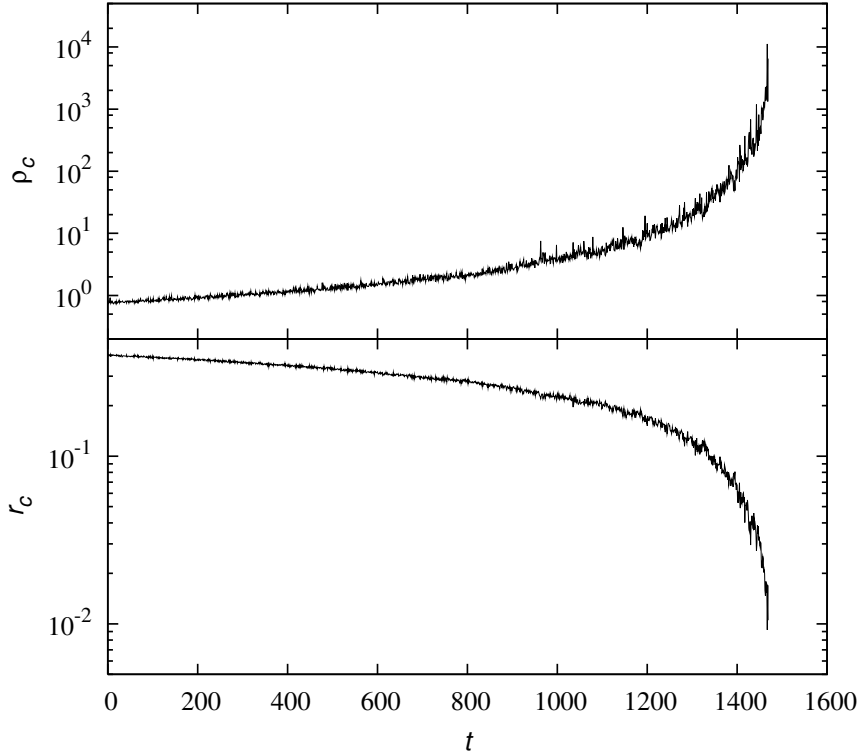}
\end{center}
\caption[Core density and core radius]{
	The core density (top) and the core radius (bottom)
	for $N$-body unit time.	
}
\label{fig:core}
\end{figure}

\subsection{Model}

The problem we study here is the evolution of a two-component star
cluster. There have been some works on this kind of system using
orbit-averaged Fokker-Plank models\cite{IW84}, but very few works
using $N$-body simulation.

We constructed an initial model in the dynamical equilibrium. It is a
Plummer model in the Heggie unit \cite{Heggie86}. We assign 10\% of
the total mass to "heavy" particles, which are 5 times more massive
than "light" particles. Thus, initial model contains 64111 light
particles and 1425 massive particles.  We used a softened gravitational
potential with softening parameter $\varepsilon=4/N$.

\subsection{Result}

Fig. \ref{fig:evolve} illustrates the evolution of the mass
segregation.  The heavy particles lose energy and sink toward the
center of the system. Fig. \ref{fig:core} shows the time evolution of
the size and mass density of the core calculated using the method
described in \cite{CasertanoHut85}. We can see that the core
shrinks and the evolution becomes faster as the core becomes smaller.
We are currently making detailed comparison between $N$-body result
and FP calculation result.

\subsection{Performance}

The total calculation time was 26.4 hours and the total number of
timesteps was $4.93\times10^{10}$, the total number of floating-point 
operations was $1.94\times10^{17}$
and the average performance was 2.03 Tflops.

Table \ref{tab:prof} gives the breakdown of the calculation time for
the period of $1046<t\le1047$ (simulation time). Average time per
blockstep is around 700 $\mu$s.  As we described in the previous
section, four complex MPI operations are performed in single
blockstep. The fact that the communication took fairly small time is
due to the very good performance of MPI library and communication
hardware of Cray XD1.

Even with this high-performance MPI library of Cray XD1, without our
2D algorithm, the communication time would increase by at least a factor
of 20, and performance would drop to less than 500 Gflops.

\begin{table}
\begin{center}
\begin{tabular}{l l}
\hline
Performance & 2.37 Gflops\\
Wall-clock  time   & 54.7 s \\
Total number of timesteps   & $3.26\times10^7$ \\
Total number of block timesteps   & $8.07\times10^4$ \\
Avg. calculation  time per blockstep  & 490 $\mu$s \\
Avg. communication time per blockstep   & 179 $\mu$s \\
Avg. $N_b$     & 404 \\
\hline
\end{tabular}
\end{center}
\caption{ Breakdown of the calculation time.}

\label{tab:prof}
\end{table}

\subsection{Scalability}

\begin{figure}
\begin{center}
  \includegraphics[height=0.75\hsize, angle=270]{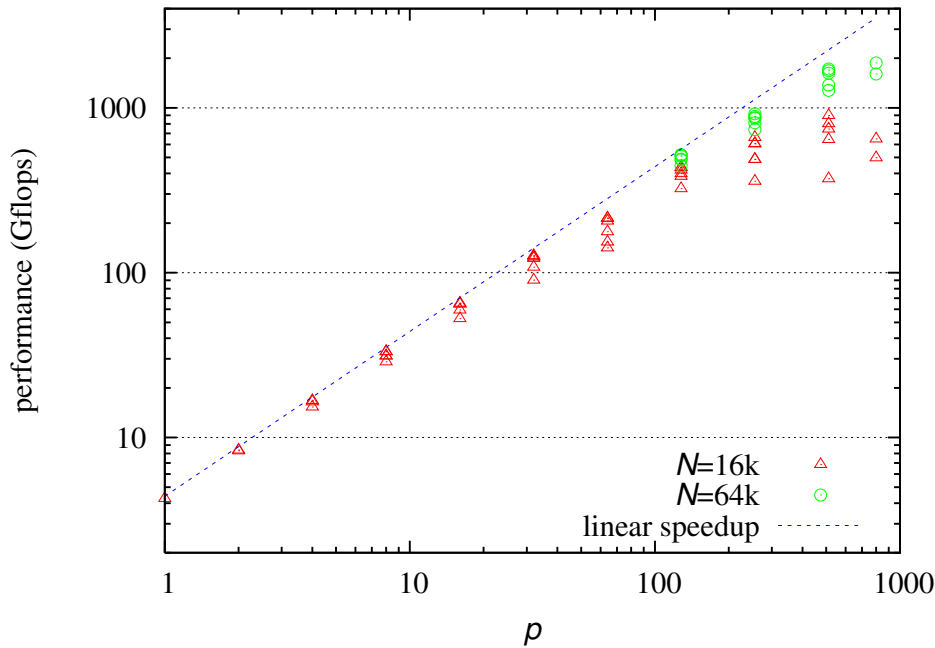}
\caption{
	Performance in Gflops as the function of number of processors
for  $N$=16k (triangle) and $N$=64k (circle).
	The dotted line indicates the ideal linear speedup.
 }
\label{fig:scale}
\end{center}
\end{figure}

Fig. \ref{fig:scale} shows the performance of our parallel $N$-body
code on a Cray XD1 with up to 800 (400 dual-core) processors, for
$N$=16K and 64K. Multiple symbols for the same values of $N$ and $p$
are results for different processor geometries. In the case of
$N$=64k, parallel efficiency is better than 80\% for up to 512
processors. Even for 16k particles, efficiency is better than 70\% for
up to 256 processors, and we achieved the performance of almost one
Tflops for 16k particles with 512 processors.

To put our performance result into perspective, we summarize some of
the previous works. Dorband et al.\cite{DHM03} obtained around 6
Gflops for $N=16{\rm k}$ on a 128-processor Cray T3E-600.  This is
around 8\% of the theoretical peak of the machine. The best number so
far reported on GRAPE-6\cite{Grape6} for $N=16{\rm k}$ is around
130 Gflops, which is 13\% of the theoretical peak. We achieved 60\% of
the theoretical peak of a 256-processor machine for the same number of
particles.

\section{Summary}

We developed  a new highly scalable parallel algorithm
for direct $N$-body simulation with individual (block) timestep
integration method.  We implemented this new algorithm on a Cray XD1 and
achieved quite good performance scaling for up to 800 processors
(the largest existing configuration of Cray XD1), for small number of
particles such as 16k or 64k. Parallel efficiency better than 80\% is
obtained for 512 processors, and sustained speed of 2.03 Tflops is
achieved for a long calculation on 800 processors.

For the first time, we demonstrated that a modern MPP with $\sim 1000$
processors can be used for astrophysical $N$-body simulations with
less than $10^5$ particles. This capability to achieve high efficiency
for small $N$ is quite important, since the total calculation cost
scales as $O(N^{3.3})$ and $N=10^5$ is the practical limit of the
number of particles we can handle with the effective speed of several
Tflops.

We thank Atsushi Kawai and Toshiyuki Fukushige of K\&F Computing
Research for their stimulating discussions.  We are grateful to Hideki
Matsumoto of SONY and staffs of Cray Japan for valuable technical
supports on the use of the XD1 system.  K.N. thanks Shigeo Mitsunari
at u10 Networks for discussions on speeding-up of the calculations
with assembly-language programming.


\begin{thebibliography}{100}
\def\pasj{PASJ}
\def\apj{ApJ}

{\parskip 0 pt

\bibitem{Aarseth63}
Aarseth,~S.~J. 1963, Monthly Notices Roy. Astron. Soc., 126, 223

\bibitem{Aarseth85}
Aarseth,~S.~J. 
1985, in Multiple Time Scales, ed. Brackbill and Cohen (Academic Press, New York), p. 377

\bibitem{CasertanoHut85} 
Casertano,~S., \& Hut,~P.\ 1985, \apj, 298, 80 

\bibitem{DHM03}
Dorband,~E.~N., Hemsendorf,~M. \& Merritt,~D. 2003,
J. Comp. Phys., 185, 484

\bibitem{Grape6A}
Fukushige,~T, Makino,~J. \& Kawai,~A. 2005, \pasj, 57, 1009

\bibitem{Gualandris2004}
Gualandris,~A., Portigies Zwart,~S. \& Tirado-Romos,~A. 2004,
submitted to IEEE Transactions on Parallel and Distributed Systems
(astro-ph/0412206)

\bibitem{Heggie86}
Heggie,~D.~C. \& Mathieu,~R.~D. 1986, 
in The Use of Supercomputer in Stellar Dynamics,
ed. P.~Hut \& S.~McMillan (New York: Springer), 233

\bibitem{HillisBarnes1987} Hillis, W.D., Barnes, J., Nature, 1987, 326, 27


\bibitem{IW84}
Inagaki,~S., Wiyanto,~P., 1984, \pasj, 36,391

\bibitem{KMI99} 
	Kokubo,~E., Makino,~J., \& Ida, S.\ 1999, 
	AAS/Division for Planetary Sciences Meeting Abstracts, 31,  

\bibitem{KokuboIda02} Kokubo,~E., \& Ida, S.\ 
2002, \apj, 581, 666 

\bibitem{HyperSys}
Lippert,~T., Ritzenh\"ofer,~G., Gl\"assner,~U, Hoeber,~H., Seyfried,~A. \& Schilling,~K. 1998
International Journal of Modern Physics C, 7, 485

\bibitem{Makino91a}
Makino,~J. 1991a, \apj , 369, 200

\bibitem{Makino91b}
Makino,~J. 1991b, \pasj, 43, 859

\bibitem{MakinoAarseth1992}
{Makino},~J. \& {Aarseth},~S.~J.
1992, \pasj,  { 44}, 141 

\bibitem{G6SC2001}
Makino,~J. \& Fukushige,~T. 2001,
In {\em The SC2001 Proceedings}, CD--ROM. Los Alamitos, IEEE Comp. Soc.

\bibitem{Makino2002}
Makino,~J. 2002, New Astronomy, 7, 373

\bibitem{Grape6}
Makino,~J., Fukushige,~T., Koga,~M., \& Namura,~K. 2003, \pasj, 55, 1163

\bibitem{McM86}
McMillan,~S.~L.~W. 1986, in The Use of Supercomputer in Stellar Dynamics,
ed. P.~Hut \& S.~McMillan (New York: Springer), 156

\bibitem{NMH06}
Nitadori,~K., Makino,~J. \&  Hut,~P. 2006, New Astronomy, submitted (astro-ph/0511062)


\bibitem{MilleniumRun} Springel, V., et al.\ 
2005b, Nat, 435, 629 

\bibitem{Grape4} Taiji, M.\ 1998, Highlights of 
Astronomy, 11, 600 

\bibitem{WSB97} 
M. S. Warren, J. K. Salmon, D. J. Becker, M. P. Goda, T. Sterling, 
and  G. S. Winckelmans, in {\it Proceedings of Supercomputing
'97}, IEEE Computer Society Press, Los Alamitos, 1997. 

}

\end{thebibliography}
\end{document}